\documentclass[aps,prl,twocolumn,graphics,epsfig,floats]{revtex4}
\usepackage{graphicx}

\newcommand{\be}{\begin{eqnarray}}
\newcommand{\ee}{\end{eqnarray}}

\def\r{\rangle}

\begin{document}

\title{Robust long-distance entanglement and a loophole-free Bell test with ions and photons}

\author{Christoph Simon$^{1,2}$ and William T. M. Irvine$^{2,1}$}
\address {$^1$ Department of Physics, University of Oxford, Parks Road, Oxford OX1
3PU, United Kingdom\\  $^2$ Department of Physics, University of
California, Santa Barbara, CA 93106}

\date{\today}

\begin{abstract} Two trapped ions that are kilometers apart can be
entangled by the joint detection of two photons, each coming from
one of the ions, in a basis of entangled states. Such a detection
is possible with linear optical elements. The use of two-photon
interference allows entanglement distribution free of
interferometric sensitivity to the path length of the photons. The
present method of creating entangled ions also opens up the
possibility of a loophole-free test of Bell's inequalities.

\end{abstract}

\maketitle

Two parties that share entanglement can use it to perform quantum
cryptography \cite{bb84,ekert} or other quantum communication
tasks, such as quantum teleportation \cite{bennett}. The creation
of entanglement between very distant locations is thus an
important goal. Owing to unavoidable transmission losses and
errors, long-distance entanglement creation is likely to require a
quantum repeater protocol, such as that of Ref. \cite{briegel}
which starts by establishing entanglement between nodes that are
separated by some basic distance, followed by entanglement
swapping and entanglement purification procedures performed at all
nodes that lie between the two desired endpoints.

Photons are the optimal systems for entanglement distribution
because they propagate fast and can preserve their coherence over
long distances. For entanglement swapping and purification, it is
advantageous to have systems that can be stored easily and between
which quantum gates can be realized efficiently. The systems used
should also have coherence times that exceed the times required
for photons to propagate over long distances. Trapped ions fulfill
all these requirements.

In a recent proposal for long-distance quantum communication
\cite{duan}, which was inspired by Ref. \cite{bose}, entanglement
between atomic ensembles is established by the detection of a
single photon that could have been emitted by either ensemble.
Here we propose a scheme where distant trapped ions are entangled
by the joint detection of {\it two} photons, one coming from each
ion. The two-photon detection, which only requires linear optical
elements and single-photon detectors, realizes a partial Bell
state analysis \cite{braunstein,densecoding}, such that for some
outcomes the two distant ions are projected into entangled states.
A practical advantage of using two-photon interference for the
creation of long-distance entanglement is that it is insensitive
to the phase accumulated by the photons on their way from the ions
to the place where they are detected.

The violation of Bell's inequality \cite{bell} by entangled
quantum systems shows that quantum physics is not compatible with
local realism. For a rigorous experimental demonstration of the
violation, firstly, the settings of the detection apparatus of two
distant observers have to be changed randomly so fast that no
information about the settings can travel from one observer to the
other at or below the speed of light during the course of each
measurement, and, secondly, the detection efficiency has to be so
high that the results of the experiment cannot be explained in
local realistic terms by systems selectively escaping detection
depending on the apparatus settings. Experiments not meeting these
two conditions are said to leave open the {\it locality} and {\it
detection} loopholes, respectively. The first one was addressed in
experiments with entangled photons \cite{aspect,weihs}, and the
second one in an experiment with trapped ions \cite{rowe}, but
closing both loopholes in a single experiment has so far remained
an elusive goal, despite several proposals \cite{loopholefree}.
Quantum state detection of ions can be performed with almost
perfect efficiency. With the proposed scheme for entanglement
creation between distant ions a final Bell experiment seems within
reach of current technology.

We will now describe the proposed protocol in more detail,
starting with the establishment of entanglement between a pair of
ions shared by two parties, Alice and Bob. Each party starts out
with an ion that contains a Lambda system of levels. The excited
state $|e\r$ can decay into two degenerate metastable states
$|s_1\r$ and $|s_2\r$ by emitting a photon into one of the two
orthogonal polarization modes $a_1$ or $a_2$. The degeneracy of
$s_1$ and $s_2$ is important to ensure indistinguishability of the
photons later on. For simplicity we will assume that the
transition probabilities for $e \rightarrow s_1$ and $e
\rightarrow s_2$ are identical as well. If this is not the case,
it can be compensated by selective attenuation of one polarization
during transmission.

The protocol starts by exciting both ions to the state $|e\r_A
|e\r_B$. Emission of a photon by each ion leads to the state \be
\frac{1}{2}[|s_1\r_A a_1^{\dagger} + |s_2\r_A
a_2^{\dagger}][|s_1\r_B b_1^{\dagger}+ |s_2\r_B
b_2^{\dagger}]|0\r, \label{initial}\ee so that each ion is
maximally entangled with the photon it has emitted. The photons
from $A$ and $B$ propagate to some intermediate location where a
partial Bell state analysis is performed. Fig. 2 shows the
well-known method for detecting the two states
$|\psi_{\pm}\r=\frac{1}{\sqrt{2}} (a_1^{\dagger}b_2^{\dagger} \pm
a_2^{\dagger}b_1^{\dagger})|0\r$ using only linear optical
elements.

\begin{figure}
\includegraphics[width=4cm]{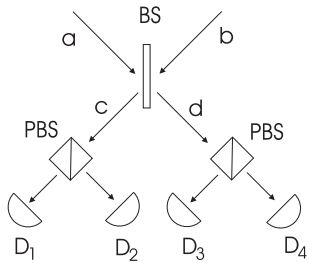}
\caption{Partial Bell state analyser as implemented in Ref.
\cite{densecoding}. One photon comes in through mode $a$, the
other through mode $b$. Only if the two photons are in the
antisymmetric Bell state $|\psi_-\r$ of Eq. (2), there will be one
photon in each output mode of the first beam splitter (BS), $c$
and $d$. Therefore a coincidence detection between $D_1$ and $D_3$
or $D_2$ and $D_4$ identifies this state. This was first pointed
out in Ref. \cite{braunstein}. If the photons are in the state
$|\psi_+\r$, they both go into $c$ or both into $d$, and are then
split by the subsequent polarizing beam splitters (PBS), because
they have orthogonal polarizations. Therefore coincidences between
$D_1$ and $D_2$ or between $D_3$ and $D_4$ signify a state
$|\psi_+\r$. For the two other Bell states $|\phi_+\r$ and
$|\phi_-\r$ both photons go to the same detector. } \label{bsa}
\end{figure}

Detection of the two photons in the state $|\psi_{\pm}\r$ will
project the two distant ions into the corresponding
$|\psi_{\pm}\r$ state, $\frac{1}{\sqrt{2}}(|s_1\r_A|s_2\r_B \pm
|s_2\r_A|s_1\r_B)$. The two remaining Bell states,
$|\phi_{\pm}\r=\frac{1}{\sqrt{2}} (a_1^{\dagger}b_1^{\dagger}\pm
a_2^{\dagger}b_2^{\dagger})|0\r$, cannot be distinguished at the
same time using only linear optical elements. The phases $\phi_A=k
L_A$ and $\phi_B=k L_B$,  where $k$ is the wave number and
$L_{A,B}$ are the path lengths, that the photons acquire on their
path from the ions to the Bell state analysis, only lead to a
multiplicative factor $e^{i(\phi_A+\phi_B)}$ in Eq.
(\ref{initial}), and thus have no effect on the entanglement. The
only condition on the path lengths is that the photon wave-packets
should overlap well in time at the detection station, since they
have to be indistinguishable in order for the Bell state analysis
to work.

The insensitivity to the phase should be contrasted with schemes
that create entanglement between two distant two-level systems,
consisting of levels $e$ and $g$, by detecting, at some
intermediate location, a single photon that could have been
emitted by either of the two systems \cite{bose,duan,plenio}. Such
a detection projects the two systems into the state
$\frac{1}{\sqrt{2}}(e^{i\phi_A}|g\r_A
|e\r_B+e^{i\phi_B}|e\r_A|g\r_B)$. The relative phase between the
two terms depends on the difference in the path lengths from each
two-level system to the intermediate location. If the relative
phase fluctuates significantly, the entanglement is destroyed.
Therefore the use of two-photon interference significantly
improves the robustness of entanglement distribution. A recent
proposal \cite{lloyd} to create entanglement between distant atoms
via the {\it absorption} of entangled photons shares this basic
advantage.

Once entangled ion states have been established between locations
separated by a certain basic distance (which is chosen depending
on the transmission losses), the length over which entangled
states are shared can be extended following the quantum repeater
protocol of Ref. \cite{briegel}. All the required quantum
operations can be performed locally at each station. Quantum gates
between ions following the proposal of Ref. \cite{cirac}, where
the ions in a trap are coupled by the phonons corresponding to
their collective motion, have recently been demonstrated
experimentally for $^{40}$Ca$^+$ ions \cite{fsknature}. A similar
scheme was previously used in Ref. \cite{turchette} to prepare an
entangled state of two $^9$Be$^+$ ions. A robust two-ion gate has
also recently been demonstrated for $^9$Be$^+$ ions
\cite{leibfried}. Establishing long-distance entanglement
following the above protocol should thus be a realistic goal.

\begin{figure}
\includegraphics[width=7cm]{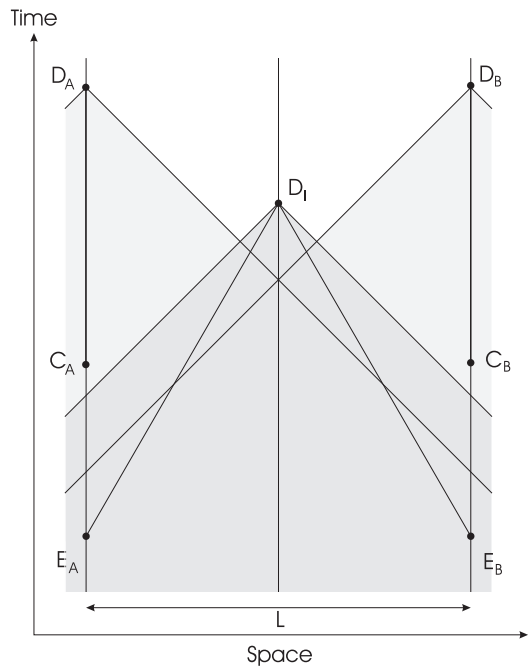}
\caption{Diagram illustrating the timing constraints for a
loophole-free Bell experiment.} \label{timing}
\end{figure}

The goal of performing a loophole-free Bell experiment puts
constraints on the timing of the various operations (Fig.
\ref{timing}). Firstly there has to be no way for the measurement
result on one side to be influenced by the choice of measurement
basis on the other side (by signals at or below $c$). This implies
that the random choice of basis on side B ($C_B$) has to lie
outside the backward lightcone of the event $D_A$, which
corresponds to the moment when the measurement result is finally
established on side $A$. Analogously, $C_A$ has to lie outside the
backward lightcone of $D_B$. This means that, for a distance $L$
between $A$ and $B$, the time that passes from the choice of basis
to the completion of detection may not be larger than $L/c$.

There is another constraint that is specific to the present
scheme. Each run of the experiment starts with the excitation of
both ions and the subsequent emission of two photons,
corresponding to the events $E_A$ and $E_B$ in Fig. \ref{timing}.
Note that the time delay between excitation and emission is
negligible in the present context. The photons propagate inside
optical fibers (with velocity $v<c$) to the intermediate location
$I$. The detection of two coincident photons in $I$ (event $D_I$)
decides that a given run is going to contribute to the data for
the Bell inequality test \cite{photodetection}. In order to avoid
a detection-type loophole for the ion pairs, the timing has to be
such that the detection or non-detection of a coincidence at $I$
can not have been influenced by the choices of measurement basis
in $A$ and $B$. That is, events $C_A$ and $C_B$ have to lie
outside the backward lightcone of $D_I$. Fig. \ref{timing} shows a
situation where this constraint is tighter than the first one.
However, since the coherence time of the ions is very long, it is
always possible to increase the time delay between excitation and
choice of basis, allowing the time between $C_A$ ($C_B$) and $D_A$
($D_B$) to approach the limit of $L/c$.

We now discuss the practical implementation of the entanglement
distribution and Bell test in more detail. To keep transmission
losses as low as possible, it would be ideal to work with photons
at the typical telecom wavelengths of 1.3 or 1.5 microns.
Unfortunately it is difficult to find suitable ionic transitions
at these wavelengths. However, suitable transitions in the visible
and near-infrared range exist, where photon losses in optical
fibers are still at an acceptable level, on the order of 1 dB/km
\cite{gisin}. This means that for a 10 km distance between Alice
and Bob, both photons will reach the intermediate station in 10\%
of the cases.  This distance corresponds to a maximum allowed time
from choice of basis to completed detection of 33 $\mu$s.

\begin{figure}
\includegraphics[width=\columnwidth]{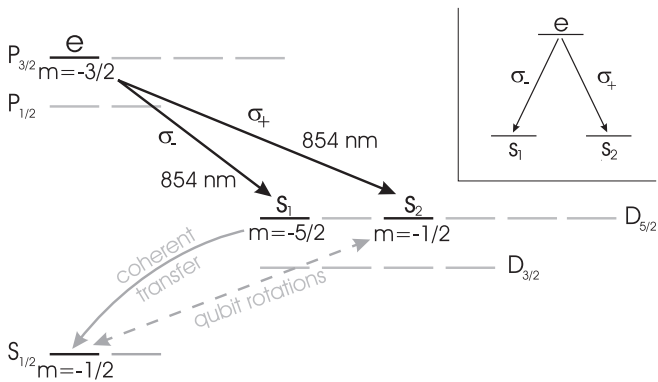}
\caption{Relevant levels of $^{40}$Ca$^+$ ions.} \label{calcium}
\end{figure}

Singly-positively charged earth alkaline ions, which have only one
electron outside a closed shell, are the most common in trapped
ion experiments. For concreteness we focus on a possible
implementation using $^{40}$Ca$^+$, whose usefulness in the
quantum information context has been demonstrated in the recent
experiments of Refs. \cite{fsknature,roos,fskjpb,mundt}. However
the use of other ions should certainly not be excluded. The
relevant levels of $^{40}$Ca$^+$ are shown in Fig. \ref{calcium}.
The metastable $D_{5/2}$ and $D_{3/2}$ states, which have
lifetimes of the order of one second, are connected to the
$P_{3/2}$ and $P_{1/2}$ levels by light with wavelengths around
860 nm. One could use one of the $P_{3/2}$ levels, say $m=-3/2$,
to serve as the excited level $e$. This state can be prepared by
first preparing the $|S_{1/2}, m=-1/2\rangle$ state by optical
pumping \cite{roos}, and then applying $\sigma_-$ light at 393 nm.
For $s_1$ and $s_2$ one could use the $m=-5/2$ and $m=-1/2$
sublevels of $D_{5/2}$, which are coupled to $e$ by $\sigma_-$ and
$\sigma_+$ light. Decay from $e$ to the $m=-3/2$ state of
$D_{5/2}$ is impossible if the photon is emitted in the direction
of the quantization axis. The rate for the $P_{3/2}\rightarrow
D_{5/2}$ transition (854 nm) is of order $0.5 \times 10^7/$s
\cite{branching}. Unwanted photons from $P_{3/2}\rightarrow
S_{1/2}$ and from $P_{3/2} \rightarrow D_{3/2}$ (850 nm) can be
avoided by spectral filtering.

It is essential to be able to detect the states of the ions in
different bases. The necessary rotations could be performed by
first coherently transferring $s_1$ ($|D_{5/2}, m=-5/2\rangle$) to
the $|S_{1/2}, m=-1/2\rangle$ state by applying a laser pulse at
729 nm as in Ref. \cite{fskjpb}, and then applying pulses of
different length to the transition between that state and $s_2$
($|D_{5/2}, m=-1/2\rangle$), exactly corresponding to single-qubit
rotations in the quantum computation scheme of Ref.
\cite{fsknature}. These transformations can be performed in a few
microseconds \cite{fskjpb}. The detection could then proceed
following Ref. \cite{fsknature}, by using a cycling transition
between the $S_{1/2}$ and $P_{1/2}$ levels (397 nm) \cite{roos}.
The $P_{1/2}\rightarrow S_{1/2}$ transition rate is of the order
of $1.3 \times 10^8$/s \cite{roos}, which determines the number of
photons that can be emitted during the cycling process. It does
not seem unrealistic to detect 1 percent of these photons, if the
collection efficiency is optimized. This would allow the detection
of 30 photons in 23 $\mu$s, more than sufficient for unambiguous
state discrimination. Sufficiently fast random switching between
different measurement bases should not be difficult to implement
using electro-optic or acousto-optic modulation
\cite{weihs,aspect}. The parameter that needs to be changed in the
present context is the length of the pulses at 729 nm.

A cavity is likely to be required for the present scheme in order
to achieve directional emission of the photons from the ions. To
determine the expected enhanced emission into the cavity mode for
the proposed $^{40}$Ca$^+$ implementation one has to take into
account not only the coupling of the $P_{3/2}\rightarrow D_{5/2}$
transition to the cavity and the cavity decay rate, but also the
fact that $P_{3/2}$ decays preferentially to $S_{1/2}$
\cite{branching}. A treatment following Ref. \cite{haroche}, but
including this loss mechanism, gives the probability $p_{cav}$ for
a photon to be emitted into the cavity mode after excitation to
$e$ as \cite{future} \be p_{cav}=\frac{4 \gamma
\Omega^2}{(\gamma+\Gamma)(\gamma \Gamma + 4 \Omega^2)}, \ee where
$\gamma=4\pi c/FL$ is the decay rate of the cavity, $F$ its
finesse, $L$ its length, $\Omega=\frac{D}{\hbar}\sqrt{\frac{hc}{2
\epsilon_0 \lambda V}}$ is the coupling constant between the
transition and the cavity mode, $D$ the dipole element, $\lambda$
the wavelength of the transition, $V$ the mode volume (which can
be made as small as $L^2\lambda/4$ for a confocal cavity with
waist $\sqrt{L\lambda/\pi}$), and $\Gamma$ is the non-cavity
related loss rate, which is equal to $1.47 \times 10^8$/s in our
case \cite{branching}. In order to maximize $p_{cav}$, $\Omega$
should be as small as possible, and $\gamma$ should be equal to $2
\Omega$. The latter condition implies an optimal cavity finesse of
$F=19000$ for our system. Coupling of a trapped ion to a 2 cm
cavity with higher finesse was recently demonstrated in Ref.
\cite{mundt}. The probability $p_{cav}$ depends strongly on the
cavity length. If one chooses $L=3$ mm, which might still be
compatible with the trap dimensions of Ref. \cite{roos}, one finds
$\gamma=9.9 \times 10^6/s$, leading to $p_{cav}=0.01$. For $L=1$
mm one would get $p_{cav}=0.06$, while $p_{cav}=0.001$ for $L=1$
cm. The above value for $\gamma$ implies that the photon
wavepackets are about 100 ns long. Such a long coherence time
makes it easy to achieve good overlap for the Bell state detection
and also makes unwanted birefringence effects in the optical
fibers negligible \cite{gisin}.

We now estimate the expected overall creation rate of two-ion
entangled pairs. Following the above timing considerations, we
assume that the ions are excited every 30 $\mu$s, corresponding to
a rate of $33000$/s. This has to be multiplied by $p_{cav}^2$, by
a factor close to 1 to describe coupling from the two cavities to
fibers, by the probabilities for both photons to reach the
intermediate station (1/10 for a 10 km distance) and to lead to a
coincident detection ($\eta^2/2$, where we assume a detection
efficiency $\eta=0.7$). Multiplying all these factors, one finds
an expected overall rate of the order of 5 pairs per minute, which
would be sufficient for performing a Bell experiment in a few
hours, and might also allow basic demonstrations of the quantum
repeater protocol. Shorter cavities would be possible for
micro-traps \cite{microtraps}, leading to greatly increased rates
(e.g. 3 pairs per second for $L=1$ mm).

In conclusion, a loophole-free test of Bell's inequalities should
be possible with just a single entanglement creation step between
two ions separated by a distance of order 10 km. Beyond that, the
present scheme allows the creation of robust long-distance
entanglement and thus significantly facilitates the realization of
quantum repeater protocols, which should allow the establishment
of entanglement over distances much larger than the absorption
length of the quantum channels.

We would like to thank F. Schmidt-Kaler, M. Arndt, N. Gisin, A.
Lamas-Linares, S. Anders, R. Blatt, D. Bouwmeester and J.-W. Pan
for answering our questions and for helpful comments. C.S. is
supported by a Marie Curie fellowship of the European Union
(HPMF-CT-2001-01205). W.I. is supported by Elsag s.p.a. (MIUR
project, grant no. 67679) and by a Broida fellowship of the
University of California at Santa Barbara.


\begin{references}
\bibitem{photodetection} A large fraction of the emitted photons will be lost on their way
to $I$. Note that this loss is not important from the point of
view of the detection loophole. It only influences the rate with
which entangled ion pairs are created.
\bibitem{branching} Ref. \cite{roos} gives the transition
probability for $P_{1/2}\rightarrow S_{1/2}$ as $1.3 \times
10^8/$s and the branching ratio of $P_{1/2}\rightarrow D_{3/2}$
versus $P_{1/2}\rightarrow S_{1/2}$ as 1:15, while Ref.
\cite{nist} gives transition probabilities $1.4 \times 10^8/$s for
$P_{1/2}\rightarrow S_{1/2}$ and $1.47 \times 10^8/$s for
$P_{3/2}\rightarrow S_{1/2}$. Following Refs. \cite{roos,sobelman}
the branching ratio for $P_{3/2}\rightarrow D_{5/2}$ versus
$P_{3/2}\rightarrow S_{1/2}$ can be estimated as 1:30, giving $0.5
\times 10^7/$s for the transition probability.
\bibitem{future} We intend to discuss this problem in more detail
in a future publication.
\end{references}
\end{document}